\documentclass[submission]{eptcs}
\providecommand{\event}{PLACES 2019} % Name of the event you are submitting to
\usepackage{underscore}
\usepackage{wrapfig}
\usepackage{amsmath}
\usepackage{amssymb, xcolor, wasysym, xspace,url}
\usepackage{prooftree}
\usepackage{tikz}
\usetikzlibrary{arrows}
\usetikzlibrary{shapes.geometric}
\usetikzlibrary{shapes,arrows,chains}
\usetikzlibrary{shapes.geometric}
\usepackage{graphicx}
\usepackage{mathtools}
\usepackage{mathptmx}
\usepackage{stmaryrd}
\usepackage{proof}
\newenvironment{proof}[1][Proof]{\begin{trivlist}
\item[\hskip \labelsep {\bfseries #1}]}{\end{trivlist}}

\newtheorem{mydef}{Definition}
\newtheorem{theorem}{Theorem}
\newtheorem{lemma}{Lemma}
\newcommand{\labelx}[1]{\label{#1}}

\newcommand{\ptilde}[1]{{\ensuremath{#1}}}
\newcommand{\kf}[1]{\textup{\textsf{#1}}\xspace}

\newcommand{\participant}[1]{\ensuremath{\mathtt{#1}}}
\newcommand{\q}{\ensuremath{\participant{q}}}
\newcommand{\p}{\ensuremath{\participant{p}}}

\newcommand{\s}{\ensuremath{s}}

\newcommand{\atw}[1]{\ensuremath{\ptilde{#1}}}

\newcommand{\mqueue}[2]{\ensuremath{#1 : #2}}
\newcommand{\emptyqueue}[1]{\mqueue{\s}{\emptyset}}

\newcommand{\G}{\ensuremath{G}}
\newcommand{\pro}[2]{\ensuremath{#1\upharpoonright #2}}

\newcommand{\T}{\ensuremath{T}}
\newcommand{\TT}{\atw{\T}}

\newcommand{\ty}{\textbf{t}}
\newcommand{\End}{\kf{end}}

\newcommand{\trule}[1]{\ensuremath{(\text{\sc{#1}})}}

\renewcommand{\bar}[1]{\overline{\,#1\,}}

\newcommand{\sss}[1]{}

\newcommand{\cloG}{\ensuremath{\mathsf{\G}}}
\newcommand{\Gli}[1]{g_{#1};G_{#1}}
\newcommand{\gli}[3]{#1_{#3};#2_{#3}}
\newcommand{\Branch}[1]{\ensuremath{\Gli{1}\times \ldots\times \Gli{#1},_{#1\in I}}}
\newcommand{\BranchG}[2]{\ensuremath{\gli{g}{#1}{1}\times \ldots \times \gli{g}{#1}{#2},_{#2 \in I}}}
\newcommand{\Brranch}[1]{\ensuremath{(\Gli{1} \times \ldots \times \Gli{#1})_{#1 \in I}}}
\newcommand{\isos}[2]{\mathrel{\mathop{\rightleftarrows}^{\mathrm{#1}}_{\mathrm{#2}}}} 
\newcommand{\gBranch}[1]{\ensuremath{g_1;g;G_1\times \ldots \times g_{#1};g;G_{#1}}}
\newcommand{\ggBranch}[3]{\ensuremath{{#1}_{#2+1};G_1\times \ldots \times {#1}_{#2+#3};G_{#3}}}

\title{Service Equivalence\\ 
via Multiparty Session Type Isomorphisms}
\author{%
  Assel Altayeva \qquad Nobuko Yoshida
  \institute{Imperial College London, UK}
}
\def\titlerunning{Service Equivalence via Multiparty Session Type Isomorphisms}
\def\authorrunning{A. Altayeva and N. Yoshida}

\begin{document}
\maketitle

\begin{abstract}

This paper addresses  a problem found within the construction of Service Oriented Architecture: the adaptation of service protocols with respect to functional redundancy and heterogeneity of global communication patterns.
We utilise the theory of Multiparty Session Types (MPST). Our approach is based upon the notion of a multiparty session type isomorphism, utilising a novel constructive realisation of service adapter code to establishing equivalence. We achieve this by employing trace semantics over a collection of local types and introducing meta abstractions over the syntax of global types.
We develop a corresponding equational theory for MPST isomorphisms.
The main motivation for this line of work is to define a type isomorphism that affords the assessment of whether two components/services are substitutables, modulo adaptation code given software components formalised as session types. 

\end{abstract}

\section{Introduction}

Multiparty session types (MPST) \cite{DBLP:conf/popl/HondaYC08} formalise multi component distributed architectures  whose semantics is necessarily given as message passing choreographies. The desired interactions are specified into a session or a \textit{global} type between participants through a series of simple syntax including interaction between two participants (composition of one send and corresponding receive), choice and recursion. Global types are then projected onto  \textit{local} types, describing communication from each participant's point of view. The theory of session types guarantees that local conformance of all participants results in of an architecture that  globally conforms to the initially specified global types.
   
   We follow the approach developed in \cite{DBLP:journals/corr/Dezani-CiancagliniPP14}. In that work, the notion of session type isomorphisms was initially explored. The main motivation for that line of work is to define the type isomorphism that would allow assessment of whether two components/services are substitutable modulo adaptation code, given component specification is considered to be a session type. This approach to isomorphism practically consists of a library of constructive combinators for witnessing this kind of equivalence. We build upon this with the intention of defining multiparty session type isomorphism combinators and study their correctness.   Comparison of two MPST is double layered: there is global syntax that is grounded in local communication semantics. 
   
   The common framework involves both {\em configurations} (collections of local types) and traces of events performed in the course of the global protocol execution as in \cite{DBLP:conf/fsttcs/DemangeonY15}. Hence syntactic change to the global protocol description might not affect local types and affords candidates for equivalent MPST. To construct combinators for  MPST isomorphisms, we employ meta abstractions over global syntax, ensuring the isomorphism preserves MPST well-formedness (projectability to local types, guaranteeing there are no orphan messages, deadlocks and that each participant has unambiguous instruction for the behaviour within the protocol). We find concurrent interactions that are independent and that do not change  the outcome if permuted. 

\paragraph{eHealth Record Example} Consider the following example, a basic eHealth record logging system. Communication is between four participants: Patient (\kf{P}), Doctor (\kf{D}), Insurance company (\kf{I}) and Hospital Record (\kf{R}). The diagram and global type protocol for it are depicted in Fig.\ref{fig:drreferral}.
The insurance company is required to give approval according to their contract conditions at each step of the treatment and at the same time hospital records have to be updated and available for further hospital and specialist systems. There are independent communications happening between Doctor and Patient, while the Insurance company makes enquiries with Hospital Records.
 
  Let us discuss the protocol in the Figure \ref{fig:drreferral}. First,  $\kf{P}\to \kf{I}:  {\langle \kf{PId,DId}\rangle}$ global type describes Patient booking appointment with the Insurance  by sending his/her identification of the type \kf{PId} and Doctor's details of \kf{DId} value type. In the second line of the global type $\cloG$ we have $\kf{D} \to	  \kf{R}:  \langle \kf{RetrRec} \rangle$, Doctor opens Patients records with the Hospital Record system, sending retrieval protocol of type \kf{RetrRec}. Next the Patient lists his "Symptoms" to the Doctor, who either prescribes a treatment or refers the Patient for further tests. At line (4), the Doctor sends his choice to the Patient and updates Patient's Hospital Record, initiating Referral or Prescription protocol with Insurance company and the Hospital.
 
 Now we observe that lines (1) and (2) follow communication between pairwise different participants, which means there is no synchronisation dependence between these two communications and they could be swapped. Another candidate for protocol transformation is the exchange between the Insurance and the Hospital Record (line (4)), triggered by the Doctor-to-Patient communication where Insurance and the Hospital Record are not aware of the order and content of the branching choices.
 
 In this work we restrict our consideration to a synchronous setting, when participants cannot start a new interaction without completing current one by waiting for the message to be received. Hence global protocol transformation will stem from the communications, which are order dependent or independent(between pairwise different participants). In the next section, we will define the formal setting for protocol transformation combinators. 
\begin{figure}
\begin{minipage}{.5\textwidth}
\scriptsize
\begin{tikzpicture}[x=1.2cm,y=1.2cm,semithick]
  \node (patient) at (1, 5) [rectangle,draw,rounded corners,minimum
  width=4em] {\normalsize P};
  \node (GP) at (2.5, 5) [rectangle,draw,rounded corners,minimum
  width=4em] {\normalsize I};
  \node (insurance) at (4, 5) [rectangle,draw,rounded corners,minimum
  width=4em] {\normalsize D};
  \node (healthrec) at (6, 5) [rectangle,draw,rounded corners,minimum
  width=4em] {\normalsize R};

  \draw [->, draw] (patient) -- (1, 1);
  \draw [->, draw] (GP) -- (2.5, 1);
  \draw [->, draw] (insurance) -- (4, 1);
  \draw [->, draw] (healthrec) -- (6, 1);
  
  \draw [->, dashed, draw] (1, 4.4) -- node [above]
  {} (2.5, 4.4);
 \draw [->, dashed, draw] (4, 4.4) -- node [above]
 {} (6, 4.4);
  
\draw [->, draw]
  (1, 4) node [circle,fill,inner sep=0pt,minimum size=3pt] {}
  -- node [above] {~~~~~~~~~~~~~~~~~~~~\kf{Symptoms}} (4, 4);

\draw [->,dotted, draw]
  (4, 3.6) node [circle,fill,inner sep=0pt,minimum size=3pt] {}
  -- node [above] {~~~~~~~~~~~~~~~~~~~~~~~\kf{Prescribe}} (1, 3.6);

  \draw [->, draw] (6, 3.2)
  -- node [above] {\qquad $\mathsf{Quote}$} (2.5, 3.2)
  node [circle,fill,inner sep=0pt,minimum size=3pt] {};

    \draw [->, draw] (6, 1.9)
  -- node [above] {\qquad $\mathsf{Quote}$} (2.5, 1.9)
  node [circle,fill,inner sep=0pt,minimum size=3pt] {};

    \draw [->, dashed, draw]
  (4, 2.9) node [circle,fill,inner sep=0pt,minimum size=3pt] {}

  -- node [above] {$\mathsf{Prescr}$} (6, 2.9);
      \draw [->, draw]
  (4, 2.4) node [circle,fill,inner sep=0pt,minimum size=3pt] {}

  -- node [above] {$\mathsf{Update}$} (6, 2.4);
  
  \draw [->,dotted, draw]
  (4, 2.2) node [circle,fill,inner sep=0pt,minimum size=3pt] {}

  -- node [above] {~~~~~~~~~~~~~~~~$\mathsf{Refer}$} (1, 2.2);
  
      \draw [->, dashed, draw]
  (4, 1.6) node [circle,fill,inner sep=0pt,minimum size=3pt] {}

  -- node [above] { $\mathsf{Ref}$} (6, 1.6);
      \draw [->, draw]
  (4, 1.3) node [circle,fill,inner sep=0pt,minimum size=3pt] {}

  -- node [above] {$\mathsf{Test}$} (6, 1.3);
   \draw [<->, draw]
  (1, 3.6) -- (0.8, 3.6) -- node[left] {choice} (0.8, 2.2) -- (1, 2.2);

\end{tikzpicture}
\end{minipage}% 
\begin{minipage}{.5\textwidth}
\small{
$%\begin{array}{@{}r@{~~}l@{}}
  \cloG= \\
\begin{array}[t]{l@{~}*{7}{l@{~}}}
   (1)& \kf{P}\to \kf{I}:  {\langle \kf{PId,DId}\rangle}; \\
  (2)&\kf{D} \to	  \kf{R}:  \langle \kf{RetrRec} \rangle; \\
  (3)& \kf{P} \to	  \kf{D}:  \langle \kf{IId,Symptoms} \rangle; \\
   (4)& \kf{D} \to \kf{P}:  \{\mathsf{Prescr:} \kf{R} \to \kf{I}:\langle \kf{Quote}\rangle;\\
 &\quad \kf{D} \to \kf{R}:\{\kf{Prescr}:\kf{D} \to \kf{R}:\langle \kf{UpRec}\rangle;\End\},\\
    & \qquad \qquad \mathsf{Ref}:\kf{R} \to \kf{I}:\langle \kf{Quote}\rangle;\\
 & \quad \kf{D} \to \kf{R}:\{\kf{Ref}:\kf{D}\to\kf{R}:\langle \kf{Test}\rangle; \End\} \}.
\end{array}$
}
\end{minipage}%

\caption{eHealth GP Visit Protocol}\labelx{fig:drreferral}
\end{figure}
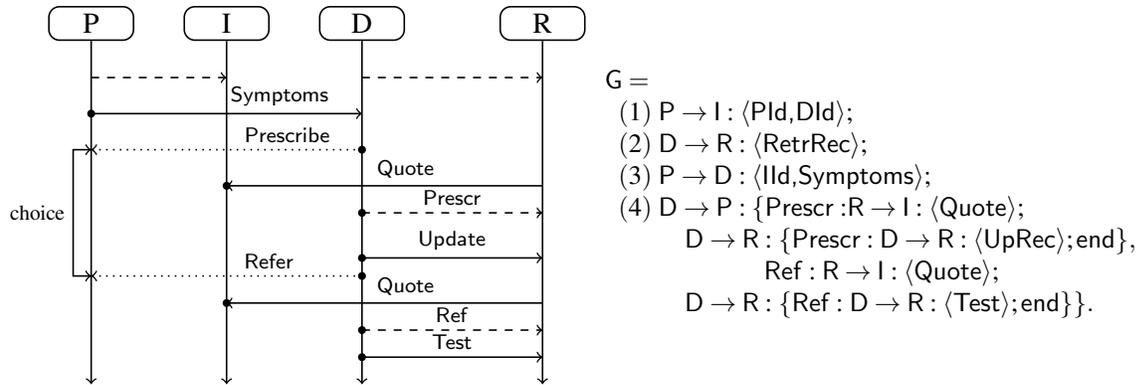

\section{Design of Multiparty Session Type Isomorphism} 

Global types define overall schemes of labelled communications between session participants. We will deconstruct the general global type syntax introduced in \cite{DBLP:conf/sfm/CoppoDPY15} to include \kf{Prefix} and \kf{Branch} subterms of the global type denoted \kf{Gtype}.  We assume base set of \textit{participants}, ranged over by $\p,\q, \kf{r},\kf{s}..$; \textit{exchange values}, ranging over boolean or natural numbers, which could also be assigned to \textit{labels} $l_1, l_2,..$ and \textit{recursion variables}, ranged over $\mathbf{t, t'..}$. 

Type formation rules for the multiparty global type abstraction are outlined in the Table ~\ref{table:Grules}.

\begin{table}[tb]
\centering
\begin{tabular}{lcr}
\hline \\
$\infer[\mbox{\scriptsize\trule{Participant}}]{\vdash \p: \mathsf{Part}}{} $ \quad
$\infer[\mbox{}]{\vdash U:\mathsf{Vtype}}{\vdash U:\mathsf{Bool}} \qquad
\infer[\mbox{\scriptsize\trule{Exchange Values}}]{\vdash U:\mathsf{Vtype}}{\vdash U:\mathsf{Nat}}$ \qquad $\infer[\mbox{\scriptsize\trule{Labels}}]{\vdash U:\mathsf{Label}}{\vdash U:\mathsf{Vtype}}$\\
\\
$\infer[\mbox{\scriptsize\trule{Global Prefix}}]{\vdash \p\to \p':\langle U \rangle :\mathsf{Prefix}}{\vdash \p,\p':\mathsf{Part} & \vdash U:\mathsf{VType}	
}$
\\
\\
$\infer[\mbox{}]{\vdash \mathsf{end}:\mathsf{Gtype}}{} \qquad \infer[\mbox{}]{\vdash \mathbf{t}:\mathsf{Gtype}}{}$ \qquad

$\infer[\mbox{\scriptsize\trule{Global type}}]{\vdash g;G:\mathsf{Gtype}}{
	\vdash g:\mathsf{Prefix} & \vdash G:\mathsf{Gtype}
} $
\\
\\
$\infer[\mbox{\scriptsize\trule{Branching}}]{\vdash \Branch{i}:\mathsf{Gtype}}{
\vdash \p,\p':\mathsf{Part}& \vdash l_i:\mathsf{Label}& \vdash G_i:\mathsf{Gtype}&	\vdash g_i:=\p \to \p':l_i: \mathsf{Prefix} & i\in I
	}
 $\\
 \\
 
 $\infer[\mbox{\scriptsize\trule{Recursion}}]{\vdash \mu \mathbf{t}.G:\mathsf{Gtype}}{
	\vdash \mathbf{t}:\mathsf{Gtype} & \vdash G:\mathsf{Gtype}
} 
$\\[1mm]
\hline

\end{tabular}
\caption{Formation Rules for Global Types}
\label{table:Grules}

\end{table}

\paragraph{Global Types} We define the syntax for MPST in the Definition \ref{def:MPSTsyntax}, introducing prefix terms for pairwise communications and predicates $\kf{inp}$ and $\kf{out}$ over it that distinguish inputting and outputting participants, which will be useful for developing global and local type trace based understanding of equivalence between two MPST.
\begin{mydef}{(Multiparty Session Types)}. Given participants $\p,\q..$, types of exchanged messages $U\in\{\mathsf{Bool, Int}\}$ and labels $l_1,...,l_n$, the grammar of global types $(G,G'..)$ is defined as:\\[1mm]
$G \quad::=\quad g;G\quad | \quad \Branch{k}\quad|
          \quad  \mathbf{t} \quad |\quad \mu \mathbf{t}.G\quad | \quad \End$\\[1mm]
$~g\quad::= \quad \p\to \q:\langle U \rangle ~~~\qquad$ $ g_i\quad::=\quad \p \to \q:l_i,\forall i \in I$\\[1mm]
$\kf{inp}(g):= \q$, $\kf{out}(g) := \p$ with 
$\kf{pid}(g)=\{\p,\q\}$; and 
$\kf{inp}(g_i):= \q$, $\kf{out}(g_i) := \p$ with $\forall i\in I.\kf{pid}(g_i)=\{\p,\q\}$.
\\[1mm]
The corresponding local session types syntax is as follows:\\[1mm]
$\T\enskip::= \enskip \kf{inp}(g)!\langle U\rangle;\T\enskip|\enskip \kf{out}(g)?\langle U\rangle;\T \enskip|
\enskip \kf{inp}(g_i)\oplus\{l_i:\T_i\} \enskip| \enskip\kf{out}(g_i) \& \{l_i:\T_i\} \enskip|
 \enskip\ty \enskip | \enskip\mu\ty. \T\enskip| \enskip \End$
 \labelx{def:MPSTsyntax}
\end{mydef} 

We will redefine the standard notion of the projection from global to
local types. Mergeability $\bowtie$ is the smallest equivalence over
local types closed under all contexts and the mergeability rule. If
$T_1 \bowtie T_2$  holds then branch merging is well-defined with a
partial commutative operator $\sqcup$. See
\cite{DBLP:conf/icalp/DenielouY13} for the full definition. 

\begin{mydef}{(Global Type Projection)} Projection ($\pro\G\ \q$) of a global type $G$ onto a participant $\q$  is defined
by induction on $\G$. Let $g=\p \to\p':\langle U\rangle$ and $g_i=\p \to \p':l_{i, \forall i\in I}$: 
\begin{quote}
{\small $\pro{(g;\G')} \q=\begin{cases}
  \kf{inp}(g)!\langle U\rangle;(\pro{\G'}\q)  & \text{if }\quad\q=\kf{out(g)} \\
  \kf{out(g)}?\langle U \rangle;(\pro{\G'}\q) & \text{if }\quad \q = \kf{inp(g)} \\
   \pro{\G'}\q & \text{otherwise}
\end{cases}$\\
$\pro{\Brranch{n}}\q=$
$
 \begin{cases}
\kf{inp}(g)\oplus\{l_i:(\pro{\G_i}\q)\}_{i\in I} & \text{if }\q=\kf{out}(g_i)_{i\in I}\\
\kf{out}(g)\&\{l_i:(\pro{\G_i}\q)\}_{i\in I}& \text{if }\q = \kf{inp}(g_i)_{i \in I} \\
   \sqcup_{i\in I}\pro{\G_i}\q  &\text{if } \q \neq \kf{pid}(g_i)_{i\in I} \quad \text{and} \\
& \forall i,j\in I.\pro{G_i}\q\bowtie\pro{G_j}\q
 \end{cases}$
   
$\pro{(\mu \ty.\G)}\q= \begin{cases}
  \mu \ty.(\pro{\G}\q)    & \text{if }\pro{\G}\q\not=\ty, \\
   \End   & \text{otherwise}.
\end{cases} \qquad \qquad\pro\ty\q=\ty\qquad\qquad\pro\End\q=
         \End$.\\[0.5mm]
}
\end{quote}
\end{mydef}

Within our example, projection onto the Insurance participant reflects his/her ignorance of the choice sent by the Doctor to the Patient, therefore Insurance is activated by the Patient sending it booking details, waits for the Quote from the Health Records and then ends: $\pro{\cloG}\kf{I}= \kf{P}?\langle \kf{PId, DId}\rangle;\kf{R}?\langle \kf{Quote}\rangle;\End$.
 
 The operational semantics for local types is defined in
Table~\ref{table:LocalLTS} with local labels set ranged over by $\ell,
\ell', ...$:
 $$\mathfrak{L}=\{ \kf{inp}(g)!m, \ \kf{out}(g)?m \quad | \quad 
m \in \{\langle U  \rangle, l\},\    
g:\kf{Prefix}, \ U:\kf{VType},\ l:\kf{Label}
\}$$ where $\kf{inp}(g)!m$ is a send action (participant $\kf{out}(g)$ is sending $m$, which could be a value or a label, to participant $\kf{inp}(g)$) and $\kf{out}(g)?m$ is a dual receive action.

 \begin{table}[!htb]

\begin{tabular}{llll}
\hline
\\ 
$ \mbox{[LIn]}$& $ \kf{out}(g)?\langle U  \rangle;\TT \xrightarrow{\kf{out}(g)?\langle U \rangle} \TT  $  &

  [LOut] &$\kf{inp}(g)!\langle U \rangle;\TT \xrightarrow{\kf{inp}(g)!\langle U \rangle} \TT     $  \\

  [LBra] & $\kf{out}(g) \& \{l_i:\TT_i\} \xrightarrow{\kf{out}(g)?
    l_j} \TT_j \quad (j \in I)    $ &

[LSel]&$\kf{inp}(g) \oplus \{l_i:\TT_i\} \xrightarrow{\kf{inp}(g)!
  l_j} \TT_j 
\quad (j \in I)     $ 
\\
  
[LRec]&$\TT[\mu \textbf{t}.\TT/\textbf{t}] \xrightarrow{\ell} \TT' \implies \mu \textbf{t}.T \xrightarrow{\ell} \TT', \quad \ell\in \mathfrak{L}$ 
\\ \\
\hline
\end{tabular}
\caption{Operational Semantics of Local Types}\label{table:LocalLTS}
\end{table}             
          
[LIn] is for a single receive action and its dual [LOut] for a send
action. Similarly, [LSel] is the rule for sending a label and its dual
[LBra] is for receiving a label. Rule [LRec] is the standard rule for recursions.   

Reduction rules for the global protocol are summarised in Table
\ref{table:opsem}.  This work focuses upon synchronous semantics for
the global type communication, following methodology  introduced in
\cite{DBLP:journals/corr/KouzapasY14}.  [Inter] shows reduction of the
global type when communication within prefix $g$ occurs, similarly
[SelBra] rule shows selecting of one of the branches $g_k$  and
executing it, which results in reduction of the global type to the
continuation of the selected branch $G_k$. Rules [IPerm] and [SBPerm]
show how to execute message passing between participants that are not
part of the the prefix communication. The last rule [Rec] is the
standard rule for recursions. 

\begin{table}[!htb]

  \begin{tabular}{llll}

 \multicolumn{4}{c}{ \noindent\rule{16cm}{0.4pt}}

\\
 $\mbox{[Inter]}$&$  g;G\xrightarrow{g}G$\qquad~~~~~~~~ & [SelBra]&$ \Branch{i}\xrightarrow{g_k} G_k$ \\[1mm]
$\mbox{[IPerm]}$&\begin{prooftree}
 G\xrightarrow{g'}G' \quad \kf{pid}(g)\cap \kf{pid}(g')=\emptyset
  \justifies
 g;G\xrightarrow{g'}g;G'
\end{prooftree}\qquad~~~~~~~
&$\mbox{[Rec]}$&$G[\mu \textbf{t}.G/\textbf{t}] \xrightarrow{g} G'
 \implies \mu \textbf{t}.G \xrightarrow{g} G'$ \\
 
 \multicolumn{4}{c}{  $\mbox{[SBPerm]}$  
\begin{prooftree}

  \forall i\in I,G_i\xrightarrow{g'}G'_i\quad \kf{pid}(g')\cap \kf{pid}(g_i)=\emptyset
  \justifies
 \Branch{i} \xrightarrow{g'}\BranchG{G'}{i}

\end{prooftree}}

\\

% $\mbox{[Rec]}$&
% \begin{prooftree}
% G[\mu \mathbf{\mathbf{t}}.G/\mathbf{t}]\xrightarrow{g}G'
% \justifies
% \mu \mathbf{\mathbf{t}}.G\xrightarrow{g}G'
% \end{prooftree}
         \\

 \multicolumn{4}{c}{ \noindent\rule{16cm}{0.4pt}}
 
    \end{tabular}
  \caption{Operational Semantics of Global Types}\label{table:opsem}
             \end{table}
 
 \begin{mydef}(Trace of a Global Type)
{Given global type $G$, we call the trace of a global type  a sequence of possible communication events during protocol execution:
  $$Tr(G)=\{g_1;g_2..;g_n |G \xrightarrow{g_1}..\xrightarrow{g_n} G', g_{i\in I}:\kf{Prefix}\}$$}
 \labelx{def:GTrace}
\end{mydef}

\paragraph{$\lambda$-Terms of MPST}
 In order to build isomorphism combinators we require two syntactic classes of variables: one called \kf{Prefix} for \textit{term} variables, and another one called \kf{Gtype} for global  \textit{type} variables. We define typed $\lambda$-terms on the variables of the \kf{Prefix} or \kf{Gtype} types:\\

\mbox{\scriptsize\trule{Variables}} \quad $ \kf{v} \quad:=\quad \kf{v}_g:\mathsf{Prefix} \quad |\quad \kf{v}_G:\mathsf{Gtype}$

\mbox{\scriptsize\trule{$\Lambda$-terms}} \quad $M \quad:= \quad \kf{v}\quad |\quad\lambda \kf{v}.M \quad|\quad$ if $e$ then $M$ else $M \quad |\quad$ let $\kf{v}=M$ in $M\quad | \quad MM       $

\mbox{\scriptsize\trule{Boolean expressions}}\quad $e\quad:=\quad \mathsf{true} \quad | \quad \mathsf{false} \quad|\quad \mathsf{not}(e) \quad | \quad e_1 \quad \mathsf{and} \quad e_2\quad | \quad e_1 \quad \mathsf{or} \quad e_2$
\\
\\
%The typing judgements for expressions will have the shape $\Gamma \vdash \kf{v}_G:G$, where \textit{typing environment} $\Gamma$ are mappings from variable either to global type Gtype or abstraction of single communication unit between two participants typed by Prefix type:
% $$\Gamma \quad := \quad \emptyset \quad|\quad \Gamma,\kf{v}:\mathsf{Vtype}\quad| \quad\Gamma,\kf{v}_g: \mathsf{Prefix} \quad| \quad \Gamma, \kf{v}_G: \mathsf{Gtype}$$
% 
 We work with the usual $\lambda$-calculus typing judgment rules for
 well-formation. Isomorphisms are combinators (functions) with respect
 to abstraction over global session type expressions, i.e. \kf{Prefix}
 for term variables and \kf{Gtype} for global type variables.

 We describe isomorphism in terms of invertible transformations over global types syntax as in the following definition:

\begin{mydef}(Global Type Isomorphism and Invertible Combinators)
{Two global types $G$ and $G'$ are isomorphic $G\isos{}{} G'$ iff there exist functions  $M: G\to G'$ and $N: G' \to G$, such that $M\circ N = \lambda x:G.x$ and $N \circ M=\lambda x:G'.x$. Terms $M,N$ are called invertible combinators.   }
 \end{mydef}

  Let us assume $g_j= F_j(G),j\in I$ where combinator $F_j$ produces  $j$-th prefix and $\mathsf{Tail}_i(G)=G'$ with  $j<i, i,j \in I$ and $\mathsf{Tail}_j(G)=g_{j+1};g_{j+2};...;g_i;G'$, we can write a swapping combinator:
\begin{equation}\tag{\textbf{Prefix commutativity}}
G= g_1;..;g_{i-1};g_i;..g_n;\overline{G}\isos{Swap^l_{g_i}}{Swap^r_{g_i}}  g_1;..;g_{i-2};g_{i};g_{i-1}..g_n;\overline{G} 
\end{equation}
where 
\begin{equation}
\begin{aligned}
\mathsf{Swap^l_{g_i}}(G) \triangleq \lambda G.\ \mathsf{let}\quad g_i=F_i(G)\  \mathsf{and}  \ G'=\mathsf{Tail}_{i}(G)\quad 
                        \mathsf{in}\quad \\
                        \mathsf {if \ pid}(g_i)\cap \kf{pid}(g_{i-1})=\emptyset
                    \quad\mathsf{then}\
                       \quad g_1;..;g_{i-2};g_{i};g_{i-1};G' \quad \mathsf{else} \quad G.
\end{aligned}                       
\end{equation}

and reverting combinator $\kf{Swap}^r_{g_i}$ will accordingly have a form:

\begin{equation}
\begin{aligned}
\mathsf{Swap^r_{g_i}}(G) \triangleq \lambda G.\ \mathsf{let}\quad g_i=F_i(G)\  \mathsf{and}  \ G'=\mathsf{Tail}_{i+1}(G)\quad 
                        \mathsf{in}\quad \\
                        \mathsf {if \ pid}(g_i)\cap \kf{pid}(g_{i+1})=\emptyset \quad\mathsf{then}\
                       \quad g_1;..;g_{i-1};g_{i+1};g_i;G' \quad \mathsf{else} \quad G.
\end{aligned}                       
\end{equation}

Returning to our example protocol $\cloG$ from the Fig \ref{fig:drreferral}, if we apply this combinator to swap  first two lines of independent communication we arrive at an isomorphic protocol  $\cloG_{sw}$:
  
 $$\cloG\isos{Swap^{l}}{Swap^{r}} 
\kf{D} \to	  \kf{R}:  \langle \kf{RetrRec} \rangle; \kf{P}\to \kf{I}:  {\langle \kf{PId,DId}\rangle}; 
   \kf{Tail}(\cloG)=\cloG_{sw} $$.

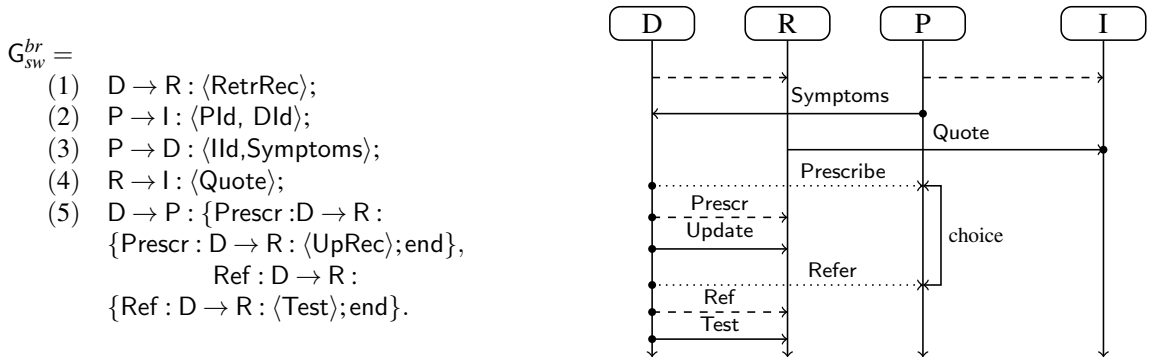
\begin{figure}
\begin{minipage}{.5\textwidth}
\small{
$%\begin{array}{@{}r@{~~}l@{}}
  \cloG^{br}_{sw}= \\
\begin{array}[t]{ll@{~~~~}*{7}{l@{~~}}}
  & (1) & \kf{D}\to \kf{R}:  {\langle \kf{RetrRec}\rangle}; \\
  & (2) & \kf{P} \to	  \kf{I}:  \langle \kf{PId, DId} \rangle; \\
 & (3) & \kf{P} \to	  \kf{D}:  \langle \kf{IId,Symptoms} \rangle; \\
 & (4)&\kf{R} \to \kf{I}:\langle \kf{Quote}\rangle;\\
  & (5) &\kf{D} \to \kf{P}:  \{\mathsf{Prescr:}\kf{D} \to \kf{R}:\\
  &&\{\kf{Prescr}:\kf{D}\to\kf{R}:\langle \kf{UpRec}\rangle;\End\},\\
  && \quad \qquad \quad\mathsf{Ref}:\kf{D} \to \kf{R}:\\
  &&\{\kf{Ref}:\kf{D}\to\kf{R}:\langle \kf{Test}\rangle; \End \}.
\end{array}$}
\end{minipage}%
\begin{minipage}{.3\textwidth}
\scriptsize
\begin{tikzpicture}[x=1.2cm,y=1.2cm,semithick]
  \node (GP) at (1, 5) [rectangle,draw,rounded corners,minimum
  width=4em] {\normalsize D};
  \node (healthrec) at (2.5, 5) [rectangle,draw,rounded corners,minimum
  width=4em] {\normalsize R};
  \node (patient) at (4, 5) [rectangle,draw,rounded corners,minimum
  width=4em] {\normalsize P};
  \node (insurance) at (6, 5) [rectangle,draw,rounded corners,minimum
  width=4em] {\normalsize I};

  \draw [->, draw] (GP) -- (1, 1.3);
  \draw [->, draw] (healthrec) -- (2.5, 1.3);
  \draw [->, draw] (patient) -- (4, 1.3);
  \draw [->, draw] (insurance) -- (6, 1.3);
  
  \draw [->, dashed, draw] (1, 4.4) -- node [above]
  {} (2.5, 4.4);
 \draw [->, dashed, draw] (4, 4.4) -- node [above]
 {} (6, 4.4);
  
\draw [->, draw]
  (4, 4) node [circle,fill,inner sep=0pt,minimum size=3pt] {}
  -- node [above] {~~~~~~~~~~~~~~~~~~~~\kf{Symptoms}} (1, 4);

\draw [->,dotted, draw]
  (1, 3.2) node [circle,fill,inner sep=0pt,minimum size=3pt] {}
  -- node [above] {~~~~~~~~~~~~~~~~~~~~~\kf{Prescribe}} (4, 3.2);

  \draw [->, draw] (2.5, 3.6)
  -- node [above] {~~~~~~$\mathsf{Quote}$} (6, 3.6)
  node [circle,fill,inner sep=0pt,minimum size=3pt] {};
  
    \draw [->, dashed,draw]
  (1, 2.85) node [circle,fill,inner sep=0pt,minimum size=3pt] {}

  -- node [above] {$\mathsf{Prescr}$} (2.5, 2.85);
  
  \draw [->, draw]
  (1, 2.5) node [circle,fill,inner sep=0pt,minimum size=3pt] {}

  -- node [above] {$\mathsf{Update}$} (2.5, 2.5);
  
  \draw [->,dotted, draw]
  (1, 2.1) node [circle,fill,inner sep=0pt,minimum size=3pt] {}

  -- node [above] {~~~~~~~~~~~~~~~~~$\mathsf{Refer}$} (4, 2.1);

      \draw [->,dashed, draw]
  (1, 1.8) node [circle,fill,inner sep=0pt,minimum size=3pt] {}

  -- node [above] {$\mathsf{Ref}$} (2.5, 1.8);
  
      \draw [->, draw]
  (1, 1.5) node [circle,fill,inner sep=0pt,minimum size=3pt] {}

  -- node [above] {$\mathsf{Test}$} (2.5, 1.5);
   \draw [<->, draw]
  (4, 3.2) -- (4.2, 3.2) -- node[right] {choice} (4.2, 2.1) -- (4, 2.1);
  
 \end{tikzpicture}
\end{minipage}

\caption{Isomorphic eHealth Protocol (branch and prefix swapping combinators applied) }
\labelx{fig:drreferral1}
\end{figure} 
  Consider communication between participants $\p$ and $\q$, where $\p$ sends choices, identified with labels $l_i, i\in I$, to proceed within each branch exchanging value $U$ between participants $\p'$ and $\q'$ and then continue as global type $G_i$  branch. Then these two communications can be swapped reflecting concurrency of the interaction:
\begin{equation}\tag{\textbf{Branching }}
\gBranch{i} \isos{Contr}{Exp}
g;(\Branch{i})
\end{equation}

\begin{equation}
\begin{aligned}
\mathsf{Contr}(G)\triangleq  \lambda g\lambda g_1 \ldots \lambda g_k\lambda  G_1 \ldots \lambda G_k.\quad \mathsf{if}\quad G=\gBranch{k} \quad \mathsf{and}\\
\mathsf{pid}(g)\cap \kf{pid}(g_i)=\emptyset,  1\leq i\leq k \quad
\mathsf{then}
        \quad g;(g_1;G_1 \times \ldots \times g_k;G_k) \quad \mathsf{else} \quad G.
 \end{aligned}
\end{equation}

The inverse of the contracting function \textsf{Contr} will be expanding one:

\begin{equation}
\begin{aligned}
\mathsf{Exp}(G) \triangleq \lambda g\lambda g_1 \ldots \lambda g_k\lambda G_1 \ldots \lambda G_k. \quad\mathsf{if} \quad G = g;(g_1;G_1 \times \ldots \times g_k;G_k) \quad \mathsf{and}\\
\mathsf{pid}(g)\cap \kf{pid}(g_i)=\emptyset, 1\leq i\leq k 
\quad \mathsf{then} \quad \gBranch{k} \quad \mathsf{else} \quad G.
\end{aligned}
\end{equation}

  The next swapping equivalence for the global type is the  analogue of the distributivity for branching within branches (indexed prefixes reflect the labels exchanged):
\begin{equation}\tag{\textbf{Branching distributivity}}
\begin{aligned}
g_1;(\ggBranch{\overline{g}}{n}{k})\times \ldots \times g_n;(\ggBranch{\overline{g}}{n}{k}) \isos{SwapBr_l}{SwapBr_r}\\
\overline{g}_{n+1};(g_1;G_1\times \ldots \times g_n;G_1)\times \ldots \times \overline{g}_{n+k};(g_1;G_k \times \ldots \times g_n;G_k),\ k\in I, n \in I \quad \mathsf{else} \quad G.
\end{aligned}
\end{equation}
In the case of the eHealth logging protocol, as we mentioned, there is an independent communication between Insurance and the Records system within the choice sent from the Doctor to the Patient. Applying contracting combinator $\mathsf{Contr}$ to the $\cloG_{sw}$ isomorphic to the $\cloG$, we arrive to another isomorphic protocol $\cloG^{br}_{sw}$ depicted in Fig.\ref{fig:drreferral1}. Projections  of this transformation $\cloG^{br}_{sw}$  of the original  protocol $\cloG$ onto the Patient, Doctor and Insurance participant are exactly the same. An interesting case arises when looking at the local type for the Hospital Record: syntactically projections on this participant of the global types from the Fig.\ref{fig:drreferral} and Fig.\ref{fig:drreferral1}  are different. However trace sets of the projections are equivalent $Tr(\pro{\cloG}\kf{R}) = Tr(\pro{\cloG^{br}_{sw}}\kf{R})$. 
\section{Semantics of Multiparty Session Type Isomorphism}
The common framework to describe session networks is to study configurations. Configurations are collections of \textit{local types} corresponding to remaining expected actions for all participants. We follow notation introduced in \cite{DBLP:conf/fsttcs/DemangeonY15} to compare sets of languages of local message traces for associated global types. 
       \begin{mydef}(Configuration Traces)
{A configuration trace $\sigma$} is a mapping from participants to
a sequence of labels of local types, i.e.~$\sigma(\kf{r})=\ell_1...\ell_n$ where $\ell_i \in \mathfrak{L}$. A participant $\kf{r}$ is in  the domain of $\sigma$ if $\sigma(\kf{r}) \neq \epsilon$ where $\epsilon$ stands for an empty sequence.  
\end{mydef} 
  
\begin{mydef}(Configurations) {Given a set of roles $\mathcal{P}$, we
    define a configuration as $\Delta=(T_\p)_{\p\in \mathcal{P}}$
    where $\TT_\p$ is a local type projected to participant $\p$
    (i.e. a local type of participant $\p$).  The synchronous transition relation between configurations is defined as}:
$$\infer[\mbox{\scriptsize\trule{Synch}}]{(T_\p)_{\p \in \mathcal{P}}\xrightarrow{\ell\cdot \overline{\ell}} (T'_\p)_{\p\in \mathcal{P}}} {T_\p \xrightarrow{\ell} T'_{\p} \quad T_\q \xrightarrow{\overline{\ell}}T'_\q \quad T_{\kf{r}}=T'_{\kf{r}} \quad \kf{r}\neq \p, \kf{r}\neq \q}$$
\end{mydef}
  
  The relation between traces and configuration is given by execution relation $\Delta\rightsquigarrow^{\sigma}_{\kf{synch}}\Delta'$:
  
  \begin{mydef}(Configuration Execution and Traces)
Configuration $\Delta$ executes trace $\sigma$ to configuration $\Delta'$ for synchronous semantics, if
 \begin{enumerate}
 \item For any configuration $\Delta$, $\Delta \rightsquigarrow^{\sigma_0}_{\kf{synch}} \Delta$,where $\forall \kf{r}: \sigma_0(\kf{r})=\epsilon$
 \item For any configurations $\Delta,\Delta_1,\Delta_2$ any trace $\sigma$, any label $l$, if $\Delta \rightsquigarrow^{\sigma}_{\kf{synch}} \Delta_1$ and $\Delta_1 \xrightarrow{\ell\cdot \overline{\ell}}\Delta_2$ within synchronous semantics, then we define $\Delta \rightsquigarrow^{\sigma'}_{\kf{synch}} \Delta_2$ as follows:
 $$ \ell\cdot \overline{\ell}=
 \q!m\cdot \p?m,\quad \mathsf{then} \quad\sigma'(\p)=\sigma(\p).\q!m,\sigma'(\q)= \sigma(\q).\p?m\quad \mathsf{and} \quad \sigma'(\kf{r})=\sigma(\kf{r}), \kf{r} \notin \{\p,\q\}.
  $$
 \end{enumerate}
\end{mydef}

\begin{mydef}(Denotation of a Global Type and Terminated Traces).
Let us define $\delta(G)=(T_\p)_{\p \in \mathcal{P}}$ 
where $\mathcal{P}$ is a set of participants in $G$. 
We   define the denotation of global type $G$ under synchronous semantic,
    denoted $\textbf{D}(G)$, as the set of all terminated traces
    from $\delta(G)$ where 
a terminated trace from $\delta(G)$ means 
$\delta (G) \rightsquigarrow^{\sigma}_{\kf{synch}}
  \Delta$ where $\Delta \not \rightarrow$. 
\end{mydef}    

Therefore we arrive to the statement about global type isomorphism with relation to the local traces:  isomorphic global types will have the same sets of traces.
 We can show for the three isomorphisms (Commutativity, Branching and Branching distributivity), that two isomorphic types will have equal trace sets, i.e. executing the same action (communication between two participants) on each isomorphic global type will result in equivalent up to defined isomorphism global types.
 
%  \begin{lemma} 
% Given $G_1 \isos{}{} G_2$  corresponding trace sets of these global types will be equal $Tr(G_1) =Tr(G_2)$ 
% \end{lemma}

\begin{lemma} \label{lem:iso:trace}
If $G_1 \isos{}{} G_2$, then $Tr(G_1) =Tr(G_2)$. 
\end{lemma}

We prove Lemma \ref{lem:iso:trace} by induction over global 
operational semantics given in Table \ref{table:opsem} for the three
isomorphism rules. The proof is given in Appendix~\ref{app:lem:iso:trace}.  

% To define equational theory $Th_{MSP}$ for the theory of global type isomorphisms, including commutativity of prefix swapping, branching and branching distributivity we prove Theorem 1 by induction on the operational semantics for global and local types.

Next Theorem~\ref{the:correspondence} shows the equivalence between
trace sets of a global type and configuration traces 
of a set of local types projected from that global type.  
% \cite[Theorem 3.1]{DBLP:conf/icalp/DenielouY13} as the sound and complete
% correspondence of the global type and local types semantics.
Let us denote the trace set of the configuration of the global type
by $\mathcal{T}_S(\Delta)$. 
The following theorem proves the trace
set of the global type $\G$ for the synchronous semantic, $Tr(\G)$,
is equivalent to $\mathcal{T}_S(\Delta)$. Below the equivalence 
relation $\equiv$ is defined by identifying 
$g=\p \to \q : m$ (the label of the global type trace)
to $\q!m\cdot \p?m$ (the labels of the configuration trace). 
The proof is given in Appendix~\ref{app:the:correspondence}.

\begin{theorem}[Equivalence between Synchronous Global Types and
  Configuration Traces]
\label{the:correspondence}
Let $\G$ be a global type with participants $\mathcal{P}$ 
and let $\Delta= (\pro{\G}\p)_{\p\in \mathcal{P}}$ 
be the local type configuration projected from $G$. 
Then $Tr(\G) \equiv \mathcal{T}_S(\Delta)$ where
$\Delta = (T_{\p})_{\p\in\mathcal{P}}$. 
\end{theorem} 

By Lemma~\ref{lem:iso:trace} and
Theorem~\ref{the:correspondence},  
Theorem~\ref{the:soundnness} concludes that the denotational semantics of
two isomorphic global types are the same. 

\begin{theorem}[Soundness]\label{the:soundnness}
Let $\G$ be a global type with participants $\mathcal{P}$. 
If $G_1 \isos{}{} G_2$, then 
$\mathcal{T}_S(\Delta_1)=
\mathcal{T}_S(\Delta_2)$ 
where $\Delta_i=(\T_{i \p})_{\p\in \mathcal{P}}$ 
with $i\in \{1,2\}$  and  $T_{i \p}=\pro{\G_i}\p$. 
Hence if $G_1 \isos{}{} G_2$, then $\textbf{D}(G_1)=\textbf{D}(G_2)$.
 \end{theorem}

We conjecture the completeness direction. 

% \begin{theorem}[Soundness]
% Equational theory $Th_{MSP}$ is a sound theory of isomorphisms for a Multiparty Session $\pi$-calculus (MSP):
% $$ \forall G,G' \quad Th_{MSP} \vdash G=G' \Rightarrow G\isos{}{} G'.$$

% \end{theorem}

% We conjecture, that equational theory $Th_{MSP}$ is a complete theory of isomorphisms for a Multiparty sessioin $\pi$-calculus, i. e. 
% whenever $\G \isos{}{} \G'$ then $\forall \G, \G'$ we have $ Th_{MSP} \vdash \G=\G'$.

\section{Related Work and Conclusions}

\paragraph{Type Isomorphism} The main theory of \emph{type isomorphisms} developed in \cite{ISOSBook} demonstrated that type theory is an effective formalism to classify software components and how type isomorphism can be practically employed to catalogue and manage behaviorally equivalent components. However, isomorphisms are often considered to be too strict in distributed settings, whose behavioural semantics is often given by means of process calculus. The need for the latter formalism of distributed component equivalence was a historical motivation for developing notions of component similarity and adaptation via bisimulation \cite{DBLP:conf/concur/Milener84}, testing equivalences \cite{DBLP:journals/corr/BernardiH15} and so on.  

In contrast, our pursuit of defining isomorphism framework for the globally governed semantic of multiparty processes is an attempt to find more flexible type level equalities. We build upon earlier work to axiomatise session type isomorphisms through behavioral adaptation. The first such attempt to investigate session type isomorphisms, following the theory of type isomorphisms \cite{DBLP:journals/mscs/Cosmo05} and finite hereditary permutations, was presented in \cite{DBLP:journals/corr/Dezani-CiancagliniPP14} and described combinators for \emph{binary session types isomorphisms} corresponding to adjacent processes.
  Interpretation of linear logic propositions as session types for communicating processes explains how type isomorphisms resulting from linear logic equivalences are realised by coercions between interface
types of session-based concurrent systems \cite{DBLP:journals/iandc/PerezCPT14}.

We extend our investigation beyond binary session types to multiparty
session types (MPST) \cite{DBLP:conf/popl/HondaYC08}.
\paragraph{Global Protocol Adaptation}
Works addressing adaptation for multiparty communications include
\cite{DBLP:journals/corr/PredaGGLM16}, \cite{DBLP:journals/soca/CoppoDV15} and \cite{DBLP:journals/fac/CastellaniDP16}                                                                                                                                                                                                                                                                                                                                                                                                                                                                                                                                                                                                                                                                                                                                                                                                                                                                                                                                                                                                                                                                                                                                                                                                                                                                                                                                                                                                                                                                                                                                                                                                                                                                                                                                                                                                                                                                                                                                                                                                                                                                                                                                                                                                                                                                                                                                                                                                                                                                                                                                                                                                                                                                                                                                                                                                                                                                                                                                                                                                                                                                                                                                                                                                                                                                                                                                                                                                                                                                                                                                                                                                                                                                                                                                                                                                                                                                                                                                                                                                                                                                                                                                                                                     .                                                                                                                                                                                                                                                                                                                                                                                                                                                                                                                                                  The paper \cite{DBLP:journals/corr/PredaGGLM16} proposes a choreographic
language for distributed applications.  Adaptation follows a
rule-based approach, in which all interactions, under all possible
changes produced by the adaptation rules, proceed as prescribed by an
abstract model. In \cite{DBLP:journals/soca/CoppoDV15} a calculus based on global types,
monitors and processes is introduced and adaptation is triggered after
the execution of the communications prescribed by a global type, in
reaction to changes of the global state. In contrast, in \cite{DBLP:journals/fac/CastellaniDP16}
adaptation is triggered by security violations, and assures access
control and secure information flow.

 \paragraph{Trace Semantics for MPST}  The first study of the expressiveness of  multiparty session types through trace semantics is 
given in \cite{DBLP:conf/fsttcs/DemangeonY15}. That work employs  sets
of languages of local message traces to compare expressiveness of
different semantics of multiparty session types based on: the presence
and nature of varied data structures (input or output queues),
flexibility of the local types, defined as a subtyping relation, and
presence of parallel sessions and interruptions. The global type
isomorphism design we offer here is straightforwardly extendable to
this semantics. 

Our future work includes the extensions to subtyping, 
sessions with interruptions, and asynchronous semantics. 
At the practical side, it is interesting to implement combinators in
functional languages such as  Haskell or OCaml taking Scribble 
\cite{DBLP:conf/tgc/YoshidaHNN13} 
as a source global protocol. 

\section{Acknowledgements}
We thank the PLACES'19 reviewers for their feedback.
The work is partially supported by
EPSRC projects
EP/K034413/1,
EP/K011715/1,
EP/L00058X/1,
EP/N027833/1,
and
EP/N028201/1.

\bibliographystyle{eptcs}
\bibliography{mybib}

\appendix
\section{Proof of Lemma~\ref{lem:iso:trace}}
\label{app:lem:iso:trace}
Recall Definition~\ref{def:GTrace} of the trace set of the global type. We will show for the three isomorphisms, that two isomorphic types will have the same trace sets, i.e. executing the same action(communication between two participants) on each isomorphic global type will result in the same or equivalent up to isomorphism global type. 
We will start with the \textit{prefix swapping} isomorphism, that
reflect the reordering of message passing for the pairwise different participants.
\begin{itemize}

\item $\underbrace{g_1;g_2;G}_{G_1} \isos{}{}\underbrace{g_2;g_1;G}_{G_2}$

\begin{proof} 
 Let us  recall operational semantics Table~\ref{table:opsem} for the Global types progress. There are three possibilities for the types $G_1$ and $G_2$ to proceed:
\begin{enumerate}
\item Execute communication $g_1$: 

Following the rule [Inter], type $G_1$ will reduce to:
$g_1;g_2:G \xrightarrow{g_1} g_2;G$. To execute the same trace on $G_2$, we apply rule [IPerm]:$g_2;g_1:G \xrightarrow{g_1} g_2;G$. Both times execution of this trace results in the same global type $g_2;G$, hence $Tr(G_1)=Tr(G_2)$.
\item Execute communication between participants in $g_2$:

By the rule [Iperm] global type $G_1$ reduces to $g_1;g_2;G\xrightarrow{g_2}g_1;G$. The same trace  on the global type $G_2$ will utilise [Inter] operational rule: $g_2;g_1:G \xrightarrow{g_2} g_1;G$. Both $G_1$ and $G_2$ reduce to the same global type $g_1;G$, which implies trace equality on this path: $Tr(G_1)=Tr(G_2)$.

\item Execute communication $g$ within global type $G$ between participants that are not involved in $g_1$ and $g_2$, reducing $G$ into $G'$:

 $$ \begin{prooftree}
 G\xrightarrow{g}G' \qquad \mathsf{pid}(g_1)\cap \kf{pid}(g_2)\cap\kf{pid}(g)=\emptyset
  \justifies
 G_1=g_1;g_2;G\xrightarrow{g}g_1;g_2;G'\qquad G_2=g_2;g_1;G\xrightarrow{g}g_2;g_1;G'
\using[\kf{IPerm}]
\end{prooftree}$$ We arrive to the global types, equivalent up to
prefix swapping isomorphism,  by induction step  on $g_1;g_2;G'$ 
and $g_2;g_1;G'$, therefore $Tr(G_1)=Tr(G_2)$.

\end{enumerate}

Thus, these two suspect isomorphic global types $G_1$ and $G_2$ have the same trace sets and by our definition of the global multiparty session type isomorphism are indeed isomorphic.
$\blacksquare$
 \end{proof}
 
 \item $\underbrace{\gBranch{i}}_{\bar{G_1}} \isos{}{}\underbrace{g;(\Branch{i})}_{\bar{G_2}}$
\begin{proof}
Following operational semantic for the global types, we distinguish three cases of the trace execution:
\begin{enumerate}
\item Execute communication  between participants offering branching selection in the $k$-th branch: $\gBranch{i} \xrightarrow{g_k} g;G_k$ by the rule [SelBra]. The same trace on the global type $\overline{G_2}$ will require application of the [IPerm] rule:
 $\begin{prooftree} 
\Branch{i} \xrightarrow{g_k}G_k\quad [\kf{SelBra}]
\justifies
g;(\Branch{i}) \xrightarrow{g_k} g;G_k
\using [\kf{IPerm}] 
 \end{prooftree}$
\item Executing communication within prefix $g$ 
$$\begin{prooftree}
g;G_k \xrightarrow{g} G_k, \quad \kf{pid}(g)\cap \kf{pid}(g_i)=\emptyset , ~ k\in I \quad \kf{[Inter]}
\justifies
\gBranch{i} \xrightarrow{g} \Branch{i}
\using[\kf{SBPerm}]
\end{prooftree}
$$
while $\overline{G_2}\xrightarrow{g} G_0$ when applying rule [Inter].
\item Executing communication $\overline{g}$ that moves global types $G_i$ into corresponding $\overline{G_i}$ is similar to the case 2 above.
\end{enumerate}
\end{proof} 
\end{itemize}

\section{Proof of the Theorem~\ref{the:correspondence}}
\label{app:the:correspondence}
\begin{proof}
By induction on reduction of the global type LTS we show that if $\delta(G) \rightsquigarrow_{\kf{synch}}^{\sigma} \delta(G')$, 
then $\mathcal{T}(\Delta) \rightsquigarrow
_{\kf{synch}}^{\sigma}\mathcal{T}(\Delta')$, i.e.~if the trace $\sigma$ is in the trace set of a global type $G$, then it is also in the trace set of the configuration corresponding to this global type.

\begin{itemize}
\item Let $G=\mathsf{end}$ .  This is a trivial case and $\delta(\mathsf{end} )\equiv \mathcal{T}(\mathsf{end})=\epsilon$.
\item Let $G=g;G_1$, where $g=\p\to \q:\langle U \rangle$ and $\mathsf{pid}(G_1)=\{\kf{r}_1,..\kf{r}_n\} =\overline{\kf{r}}$

Then configuration $\Delta$ of the global type $G$ is the set of its local projections, $\Delta = T_{\p}, T_{\q}, T_{\overline{r}}$.
There are two possibilities for the global type $G$ to proceed according to the operational semantics of LTS in Table \ref{table:opsem}:
\begin{enumerate}
\item $G\xrightarrow{g} G'$ 
Then configuration of the global type $G$ will follow the transition
relation [$\mathsf{Synch}$] rule with the trace  $\sigma=\kf{inp}(g)!\langle U \rangle \cdot \kf{out}(g)?\langle U \rangle$:
$$\Delta = \kf{inp}(g)!\langle U \rangle; T_{\kf{out}(g)}, \kf{out}(g)?\langle U \rangle; T_{\kf{inp}(g)},
T_{\overline{r}} \xrightarrow{\kf{inp}(g)!\langle U \rangle \cdot \kf{out}(g)?\langle U \rangle} 
T_{\kf{out}(g)}, T_{\kf{inp}(g)}, T_{\overline{r}}= \Delta'$$ 

At the same time $\delta_S(G'): T_{\kf{out}(g)}, T_{\kf{inp}(g)},T_{\overline{r}}$.

\item $G\xrightarrow{g'} G'$, where $g'=\kf{r}\to \kf{s}:\langle U'
  \rangle$ and $\kf{empty}_S(g,g')$, the trace we execute in this case
  will be $\sigma=\kf{inp}(g')!\langle U' \rangle \cdot \kf{out}(g')?\langle U' \rangle$
\begin{multline}
$$\Delta = \kf{inp}(g)!\langle U' \rangle ;T_{\kf{out}(g)},
\kf{out}(g)?\langle U' \rangle ; T_{\kf{inp}(g)},\kf{inp}(g')!\langle
U' \rangle ; T_{\kf{out}(g')}, \kf{out}(g')?\langle U' \rangle ;T_{\kf{inp}(g')} T_{\overline{r}} \rightarrow \\
\xrightarrow{\kf{inp}(g')!\langle U' \rangle \cdot
  \kf{out}(g')?\langle U' \rangle} \kf{inp}(g)!\langle U' \rangle ;T_{\kf{out}(g)}, \kf{out}(g)?\langle U' \rangle; T_{\kf{inp}(g)}, T_{\kf{out}(g')}, T_{\kf{inp}(g')}, T_{\overline{r}}= \Delta'$$ 
\end{multline}

Transition of the configuration $\Delta$ reduces to the configuration $\Delta'$, meanwhile trace of the reduced global type $\delta(G')= \Delta'$
\end{enumerate}
\item Let $G=\Branch{k}$,  where $g_i=\p\to\q:l_i,i\in I$ and $\kf{pid}(G_i)=\overline{\kf{r}}_i$.
There are two cases depending on the transition rule used for the global type operational semantics:
\begin{enumerate}
\item $G\xrightarrow{g_i} G_i$ following $[\kf{SelBra}]$ rule.
Corresponding global type configuration $\Delta$ will be executing trace $\sigma=\kf{inp}(g_i)!l_i\cdot \kf{out}(g_i)?l_i$.
\item $G\xrightarrow{g'_i} G'$ using $[\kf{SBPerm}]$ rule.
Corresponding configuration will be executing trace $\sigma=\kf{inp}(g')!l'_i\cdot \kf{out}(g')?l'_i$.
In both cases configuration trace of the branching global type will be coinciding  with reduced global type configuration.

\end{enumerate}

\end{itemize}

$\blacksquare$
\end{proof}

\end{document}